\documentclass[aps,prb,twocolumn,floatfix,superscriptaddress,showpacs,footinbib]{revtex4}
\usepackage{amsmath,amssymb,natbib,bm,graphicx,url,epsfig,psfrag}
\usepackage[ansinew]{inputenc}
\usepackage{color}

\begin{document}

\title{Cotunneling and non-equilibrium magnetization in magnetic
molecular monolayers}

\author{Florian Elste}
\email{felste@physik.fu-berlin.de}
\affiliation{Institut f\"ur Theoretische Physik, Freie Universit\"at Berlin,
Arnimallee 14, 14195 Berlin, Germany}
\author{Carsten Timm}
\email{ctimm@ku.edu}
\affiliation{Department of Physics and Astronomy, University of Kansas,
Lawrence, Kansas 66045, USA}

\date{November 4, 2006}

\begin{abstract}
Transport and non-equilibrium magnetization in monolayers of magnetic
molecules subject to a bias voltage are considered. We apply a
master-equation approach going beyond the sequential-tunneling approximation
to study the Coulomb-blockade regime. While the current is very small in
this case, the magnetization shows changes of the order of the saturation
magnetization for small variations of the bias voltage. Inelastic cotunneling
processes manifest themselves as differential-conductance steps, which
are accompanied by \emph{much larger} changes in the magnetization. In
addition, the magnetization in the Coulomb-blockade regime exhibits strong
signatures  of sequential tunneling processes \emph{de}-exciting molecular
states populated by inelastic cotunneling. We also consider the case of a
single molecule, finding that cotunneling processes lead to the occurrence of
\emph{magnetic sidebands} below the Coulomb-blockade threshold. In the context
of molecular electronics, we study how additional spin relaxation suppresses
the fine structure in transport and magnetization.
\end{abstract}


\maketitle

\section{Introduction}

The idea of \emph{molecular spintronics} consists of integrating the promising
concepts of molecular electronics and
spintronics.\cite{Emberly,Nitzan,Xue,Sanvito,Wolf,Zutic} A particularly
interesting aspect of molecular electronics, besides the prospect of further
miniaturization, is the possibility to use chemical synthesis for the
fabrication of device components. This bottom-up
process would start from relatively
simply molecules and be massively parallel. In this
context, spintronics is discussed in relation to magnetic memory\cite{Joachim}
and quantum computation.\cite{Har02} Both ideas rely on \emph{magnetic}
molecules.\cite{Blundell} Partly for this reason electronic transport
through magnetic molecules has recently received a lot of
attention.\cite{JPark,Liang,PaF05,Elste,Romeike,Romeike2,Heersche,Jo,Timm,Elste2,LeM06,DGR06,RWH06,MiB06,GoL06}

Experimental research has focused on the fine struc\-ture of the
Cou\-lomb-blockade peaks\cite{JPark,Liang,Heersche,Jo} and on Kondo
correlations in single-molecule transistors.\cite{JPark,Liang,PCH06}
Furthermore, novel spin-blockade mechanisms and negative differential
conductance have been observed.\cite{Heersche,Jo} These findings have also
stimulated theoretical work, which mostly employs the sequential-tunneling
approximation.
\cite{Elste,Romeike2,Timm,Elste2,LeM06,DGR06,MiB06,Golovach,Kim} Like
artificial quantum dots, a molecular junction is in the Coulomb-blockade regime
at sufficiently small bias voltage, except at crossing points where two states
with electron numbers differing by unity become degenerate. Due to the
discreteness of molecular many-particle energies for weak coupling to the
leads, there are typically no molecular transition energies in the window
between the chemical potentials of the leads for small bias. In this regime the
very small tunneling current is not correctly described by the
sequential-tunneling approximation. It is instead dom\-i\-na\-ted by
\emph{co\-tun\-nel\-ing}, which appears in fourth order in the perturbation
expansion in the tunneling amplitude. How\-e\-ver, despite its experimental
observation,\cite{Jo} transport through magnetic molecules in this regime has
been little studied.\cite{GoL06}

We study magnetic molecules under a bias voltage in the Coulomb-blockade
regime. Our main result is that while any features in the differential
conductance are very small due to the suppression of the current, there are
\emph{large} changes in the average magnetic moments of the molecules with bias
voltage and applied field. The measurement of magnetic moments of
sub-monolayers of molecules has been demonstrated 20 years ago.\cite{Zomack}
Even the detection of the spin of a single molecule may be
feasible.\cite{Durkan,Rugar} However, it is not clear how to perform such a
measurement in a molecular-junction experiment. 
One recent experiment suggests that it is possible to employ carbon nanotube 
superconducting quantum interference devices for the
detection of the switching of singe magnetic moments.\cite{Cleuziou}
We here mainly consider a \emph{monolayer} of magnetic molecules between metallic electrodes, since
the measurement of the magnetization of a thin film is expected to be easier
than of a single molecule. Various molecules form nearly
perfect monolayers on metallic substrates.\cite{Zhou}

To find the current and the non-equilibrium magnetization, we use the
master-equation formalism, treating the tunneling to the leads as a
perturbation.\cite{Blum,BrF04,Mitra,Elste,Timm,Elste2} This approach describes
the Coulomb and exchange interactions on the molecule \emph{exactly} and works
also far from equilibrium. In particular, it is not restricted to the
li\-near-res\-ponse regime of small bias voltage.

For memory applications the control of spin relaxation is crucial. Since
cotunneling and additional spin relaxation due to, e.g., dipolar and hyperfine
interactions, have similar selection rules for molecular transitions,
consistency requires to include both. We find that spin relaxation
is very effective in washing out the fine structure in the Coulomb-blockade
regime but may be used to advantage for the generation of spin-polarized
currents.

\section{Model and Methods}

\begin{figure}
\begin{center}
\includegraphics[width=6.5cm,angle=0]{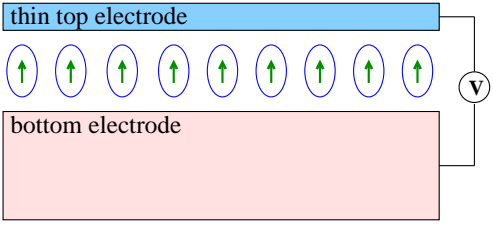}
\caption{(Color online) Sketch of the geometry.
A monolayer of magnetic molecules is adsorbed on
a metallic substrate, which serves as a bottom electrode. A thin metallic layer
is used as a top electrode.}\label{Fig1}
\end{center}
\end{figure}

For the most part, we consider a monolayer of magnetic molecules sandwiched
between two metallic electrodes, see Fig.~\ref{Fig1}. We assume that magnetic
interactions between the molecules are negligible
and that all molecules have the same spatial
orientation relative to the electrodes.\cite{Zhou}
In this case it is sufficient to
consider the properties of a single molecule.
Relaxation in the leads is assumed to be fast so that
their electron distributions can be described by equilibrium Fermi
functions. In the simplest case, transport involves tunneling
through only a single molecular level with onsite energy $\epsilon_d$ and local
Coulomb repulsion $U$. The full Hamiltonian of the system reads 
$H = H_{\text{mol}} + H_{\text{leads}} + H_{\text{t}}$, where\cite{Timm}
\begin{align}
H_{\text{mol}}\: =\: & \epsilon_d \, n_d + \frac{U}{2}n_d\left(n_d-1\right)
-J\,\mathbf{s}\cdot\mathbf{S} \nonumber \\
& {}-K_2(S^z)^2 - B\left(s^z+S^z\right) \label{Hamiltonian}
\end{align}
describes the molecular degrees of freedom,
$H_{\text{leads}} = \sum_{\alpha=\text{L},\text{R}} \sum_{\mathbf{k} \sigma}
  \epsilon_{\alpha \mathbf{k}}
  a_{\alpha \mathbf{k} \sigma}^{\dagger} a_{\alpha \mathbf{k} \sigma}$
represents the two leads $\alpha=\text{L},\text{R}$ (left, right), and
$H_{\text{t}}= \sum_{\alpha=\text{L},\text{R}} t_\alpha \sum_{\mathbf{k}\sigma}
  ( a_{\alpha \mathbf{k} \sigma}^{\dagger} c_{\sigma}
  + c_{\sigma}^{\dagger} a_{\alpha \mathbf{k} \sigma})$
describes the tunneling. The tunneling amplitudes $t_\alpha$ are chosen
real.
The operator $c_{\sigma}^{\dagger}$ creates an electron with spin
$\sigma$ on the molecule. $n_d
\equiv c^{\dagger}_{\uparrow}c_{\uparrow} +
c^{\dagger}_{\downarrow}c_{\downarrow}$ and $\mathbf{s} \equiv \sum_{\sigma
\sigma'} c^{\dagger}_{\sigma} \mbox{\boldmath$\sigma$}_{\sigma\sigma'}
c_{\sigma'} /2$  are the corresponding number and spin operator,
respectively. The parameter $J$ denotes
the exchange interaction between the electrons and a local spin
$\mathbf{S}$, where $\mathbf{S}\cdot\mathbf{S}=S(S+1)$.
We restrict ourselves to the case of easy-axis
anisotropy, $K_2>0$. For simplicity we consider 
identical g factors for $\mathbf{s}$ and $\mathbf{S}$. 
An external magnetic field $B$ is applied along the easy axis of the molecule, 
where a factor $g\mu_B$ has been absorbed into $B$. 
$a_{\alpha \mathbf{k} \sigma}^{\dagger}$ creates an electron in lead 
$\alpha$ with spin $\sigma$, momentum $\mathbf{k}$ and  energy $\epsilon_{\alpha \mathbf{k}}$.   

The leading contribution to the transition rates between molecular many-particle
states is of second order in $H_{\text{t}}$, corresponding to
sequential tunneling.
The transition rates can be obtained from Fermi's Golden Rule,\cite{BrF04}
\begin{align}
\Gamma^{nn'}_\alpha = 2\pi \sum_{\sigma} & t_{\alpha}^2 \nu_{\alpha\sigma}
\Big( f(\epsilon_m-\epsilon_n-\mu_{\alpha})|C_{nm}^{\sigma}|^2 \nonumber \\
& {}+ \left[1-f(\epsilon_n-\epsilon_m-\mu_{\alpha})\right] |C_{mn}^{\sigma}|^2 \Big).
\end{align}
Here, the eigenstates $|n\rangle$ and $|n'\rangle$ of $H_{\text{mol}}$ denote
the initial and final state of the molecule, respectively, $\nu_{\alpha\sigma}$
is the density of states per unit cell of
electrons with spin $\sigma$ in lead $\alpha$,
$f(\epsilon)$ is the Fermi function, 
$\mu_\alpha$ is the chemical potential in lead $\alpha$, where
$\mu_{\text{L}}-\mu_{\text{R}}=-eV$,
and $C^{\sigma}_{nn'}\equiv
\langle n | c_{\sigma} | n' \rangle$ is the matrix element of the electron
annihilation operator between molecular many-particle states. The typical
sequential-tunneling rate involving lead $\alpha$ and electrons with spin
$\sigma$ is given by $1/\tau_{\alpha\sigma} = 2\pi\,t_\alpha^2 \nu_{\alpha
\sigma}$ ($\hbar$ is set to unity).

To go beyond the leading order, the tunneling Hamiltonian is replaced by
the $T$ matrix,\cite{BrF04} which is self-consistently given by
\begin{equation}
T = H_{\text{t}} + H_{\text{t}}\, \frac{1}{E_i-H_0+i\eta}\,T .
\end{equation}
Here, $E_i$ is the energy of the initial state 
$|i\rangle |n\rangle$, where $|i\rangle$ refers to the equilibrium state of the left
and right leads (at different chemical potential)
and $|n\rangle$ is a molecular state. Furthermore, $H_0\equiv H_{\text{mol}} +
H_{\text{leads}}$, with the energy of the leads measured relative to
equilibrium, and $\eta$ is a positive infinitesimal
ensuring that the Green function in $T$ is retarded.
To fourth order, the transition rate from state $|i\rangle|n\rangle$ to
$|f\rangle|n'\rangle$ with an electron tunneling from lead $\alpha$ to lead
$\alpha'$ is given by
\begin{align}
\Gamma_{\alpha\alpha'}^{ni;n'f} = & 2\pi 
\left| \langle f | \langle n' | H_{\text{t}} \frac{1}{E_i-H_0+i\eta} H_{\text{t}}
  | n \rangle | i \rangle \right|^2 \nonumber \\
& {}\times  \delta(E_f-E_i) .
\end{align}
The energies of the initial state $|n\rangle |i\rangle$ 
and final state $|n'\rangle |f\rangle = |n'\rangle a^{\dagger}_{\alpha' \mathbf{k}'
\sigma'} a_{\alpha \mathbf{k} \sigma} |i\rangle$
are denoted by $E_i$ and $E_f$, respectively.
We restrict ourselves to the case of infinite $U$,
i.e., double occupancy of the molecule is forbidden.
Inserting $H_{\text{t}}$ and summing over final lead states yields
\begin{align}
\Gamma_{\alpha\alpha'}^{nn',00} ~=~& 2\pi t_{\alpha}^2 t_{\alpha'}^2
\sum_{\sigma\sigma'} \nu_{\alpha\sigma} \nu_{\alpha'\sigma'} \nonumber \\
& \times \int d\epsilon \left| \sum_{n''} \frac{C^{\sigma'}_{n'n''}C^{\sigma*}_{nn''}}{\epsilon+\epsilon_{n}-\epsilon_{n''}+i\eta} \right|^2 \nonumber \\
& \times f(\epsilon-\mu_\alpha)~[1-f(\epsilon+\epsilon_n-\epsilon_{n'}-\mu_{\alpha'})], \label{cot-rates00} \\
\Gamma_{\alpha\alpha'}^{nn',11} ~=~& 2\pi t_{\alpha}^2 t_{\alpha'}^2
\sum_{\sigma\sigma'} \nu_{\alpha\sigma} \nu_{\alpha'\sigma'} \nonumber \\
& \times \int d\epsilon \left| \sum_{n''} \frac{C^{\sigma'}_{n''n}C^{\sigma*}_{n''n'}}{-\epsilon+\epsilon_{n'}-\epsilon_{n''}+i\eta} \right|^2 \nonumber \\
& \times f(\epsilon-\mu_\alpha)~[1-f(\epsilon+\epsilon_n-\epsilon_{n'}-\mu_{\alpha'})], \label{cot-rates11}
\end{align}
where $\Gamma_{\alpha\alpha'}^{nn',00}$ ($\Gamma_{\alpha\alpha'}^{nn',11}$)
denotes the cotunneling rate describing virtual transitions between two empty 
(singly occupied) molecular states. Here, we have assumed that the density of
states in the leads is independent of energy. To the same order in
$H_{\text{t}}$, one also obtains processes changing the electron number by $\pm
2$. For $U\to\infty$ these pair-tunneling processes\cite{KRO06} are suppressed.
Note that Eqs.~(\ref{cot-rates00}) and (\ref{cot-rates11}) contain both elastic
and inelastic cotunneling.

Since the above expressions diverge due to second-order poles from the energy
denominators, the cotunneling rates cannot be evaluated
directly.\cite{Ave94,TuM02,KOO04,Koch2} We apply a regularization scheme
that follows Refs.~\onlinecite{TuM02,KOO04,Koch2} and is motivated by the
observation that Eqs.~(\ref{cot-rates00}) and (\ref{cot-rates11})
do not take into account
that the intermediate state obtains a finite width $\Gamma$ due to
tunneling. In our regime of weak tunneling, the width $\Gamma$ is of second
order in the tunneling amplitudes $t_\alpha$. This width is introduced into the
energy denominators, replacing $\eta$. When the cotunneling rates are expanded
in powers of $\Gamma$, it turns out that the leading term is of order
$1/\Gamma\propto 1/t_\alpha^2$. This cancels two powers of the tunneling
amplitude in Eqs.~(\ref{cot-rates00}) and (\ref{cot-rates11}) so that the
result is in fact a \emph{sequential-tunneling} contribution. Since we have
already included the full sequential-tunneling rates, this new contribution
should be dropped. We thus take the next order, $\Gamma^0$, for the cotunneling
rates.

The ad-hoc regularization of the cotunneling rate 
is not necessary in a description of cotunneling through wide quantum dots 
using non-equilibrium Green functions.\cite{Koenig} 
This approach avoids the divergences and also leads to a renormalization 
of transition energies at fourth order in the tunneling Hamiltonian. 
For a fully quantitative description of cotunneling through magnetic 
molecules it would be desirable to employ this approach, which is, however, 
made complicated by the presence of the internal spin degree of freedom. 
On the other hand, the T-matrix approach used here is found to give 
qualitatively reasonable results in comparison with cotunneling experiments\cite{Jo}
and we expect it to catch the relevant physics for the system studied here.

The sequential and cotunneling rates appear in the rate equations
for the probabilities to find the molecule in state $|n\rangle$
(we assume rapid dephasing\cite{Elste}),
\begin{align}
\frac{d P^n}{dt} = & \sum_{\alpha m}
  \big( \Gamma^{mn}_{\alpha} P^m - \Gamma^{nm}_{\alpha} P^n \big) \nonumber \\
& {}+ \sum_{\alpha\alpha'm} \big( \Gamma^{mn}_{\alpha\alpha'} P^m 
  - \Gamma^{nm}_{\alpha\alpha'} P^n \big) ,
\label{rate-equations}
\end{align}
where $\Gamma^{mn}_{\alpha\alpha'}\equiv \Gamma^{mn,00}_{\alpha\alpha'} +
\Gamma^{mn,11}_{\alpha\alpha'}$ and
$\Gamma^{mn,00}_{\alpha\alpha'}$ ($\Gamma^{mn,11}_{\alpha\alpha'}$)
is non-zero only if both $|n\rangle$ and $|m\rangle$ are empty
(singly occupied). The current through the left lead is given by
\begin{equation}
I^{\text{L}} = -e\sum_{nm}(n_n-n_m) \Gamma^{mn}_{\text{L}} P^m
- e\sum_{nm} \left( \Gamma^{mn}_{\text{L}\text{R}} - \Gamma^{mn}_{\text{R}\text{L}} \right) P^m .
\end{equation}
The \emph{steady-state} probabilities $P^m$ of molecular states
are obtained by solving Eq.~(\ref{rate-equations}) with the time derivatives
set to zero. The average magnetization in the $z$ direction
per molecule is given by
$M = \sum_n m_n P^n$, where $m_n$ denotes the 
quantum number of the $z$ component of the total spin $\textbf{s}+\textbf{S}$ in state
$|n\rangle$.

\section{Results and Discussion}

\begin{figure}[t]
\begin{center}
\includegraphics[width=8cm,angle=0]{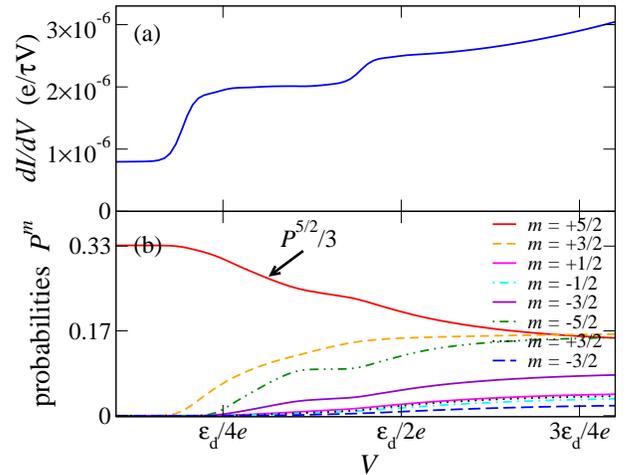}
\caption{(Color online)
(a) Differential conductance $dI/dV$ and (b) probabilities $P^n$ of
molecular many-particle states as functions of bias voltage $V$,
for low bias voltages.
The probability $P^{5/2}$ of the ground state has been scaled by a factor
$1/3$.
Here, we assume $S=2$, $J=K_2=5\,\mathrm{meV}$, $\epsilon_d=10\,J$,
$B=2\,\mathrm{meV}$, and
$T=0.3\,\mathrm{meV}$.\protect\cite{footnote.values}}\label{Fig2}
\end{center}
\end{figure}

\begin{figure}[t]
\begin{center}
\includegraphics[width=7.5cm]{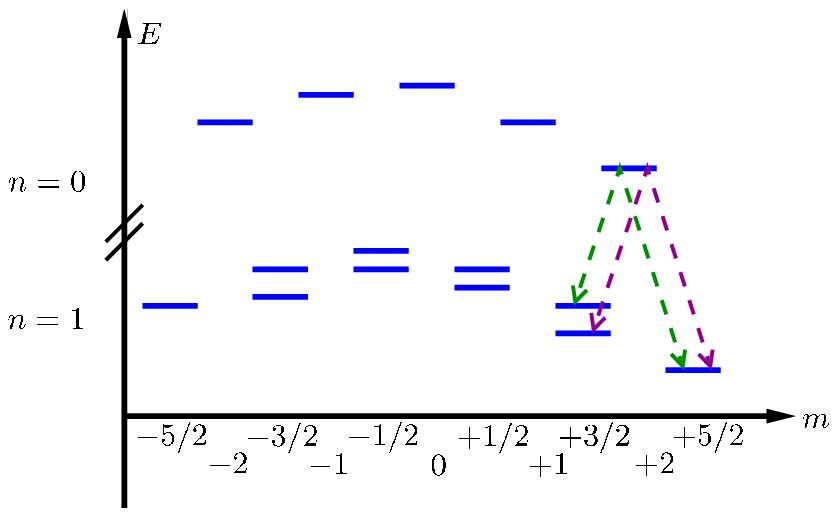}
\caption{(Color online) Level scheme
showing the energies of molecular states as a function of magnetic quantum
number $m$ for electron numbers $n=0,1$.\protect\cite{footnote.values}
The dashed double arrows signify
inelastic cotunneling between the ground state with $m=5/2$ and the two
states with $m=3/2$, involving virtual occupation of the state with $n=0$ and
$m=2$.
While sequential tunneling requires a change of the electron number by
$\Delta n=\pm 1$ and
of the magnetic quantum number by $\Delta m = \pm 1/2$,
cotunneling processes obey the selection rules $\Delta n=0$,
$\Delta m = 0,\pm 1$.}\label{Fig3}
\end{center}
\end{figure}

We start by discussing the results obtained for the differential conductance
$dI/dV$ at low bias voltages. If the system is in the Coulomb-blockade regime,
sequential tunneling is thermally suppressed and transport is dominated by
cotunneling. The magnitude of the current is then small. The conductance
at zero bias voltage is finite, see Fig.~\ref{Fig2}(a), due to \emph{elastic}
cotunneling. The cotunneling rates are proportional to the bias voltage, if
the molecular level is far from the chemical potentials, leading to ohmic
behavior. The rounded steps in $dI/dV$ correspond to the onset of additional
\textit{inelastic} cotunneling processes. Selection rules for the spin quantum
number require $\Delta m = 0,\pm 1$. For the parameters chosen in
Fig.~\ref{Fig2}, the ground state has electron number $n=1$ and maximum spin,
$m=5/2$. Inelastic cotunneling processes corresponding to the two steps involve
the two different final states with $n=1$, $m=3/2$ and virtual occupation of
the state with $n=0$, $m=2$, as illustrated in Fig.~\ref{Fig3}. Further steps
in $dI/dV$ are not observed, since the corresponding  inelastic cotunneling
transitions have smaller energy differences
between initial and final states and are therefore activated
immediately when the probability of the initial state becomes significant.

Cotunneling steps and sequential tunneling peaks show fundamentally different
dependences on the onsite energy $\epsilon_d$. For \emph{single-molecule}
junctions it is possible to change $\epsilon_d$ by applying a gate
voltage, e.g., in molecular-junction experiments. 
For monolayers one does not have this opportunity. We come back to this point below. While the bias
voltages at which sequential tunneling peaks occur shift linearly with
$\epsilon_d$, the positions of cotunneling steps remain unaffected. This
follows directly from evaluating Eqs.~(\ref{cot-rates00})--(\ref{cot-rates11})
in the limit of large $\epsilon_d$.\cite{BrF04} For magnetic molecules, the
position of the cotunneling steps shifts linearly as a function of the external
magnetic field due to the Zeeman effect, as observed for
$\mathrm{Mn_{12}}$.\cite{Jo}

While $dI/dV$ represents the change of the \emph{very small} current with bias
voltage in the cotunneling regime, the change of the probabilities $P^n$
of molecular states with bias voltage is of order unity, as shown in
Fig.~\ref{Fig2}(b). The probability of the lowest-energy state with $m=5/2$
decreases, whereas the probabilities of other states increase. Cotunneling
enables transitions between molecular states with the same electron number but
with magnetic quantum numbers differing by $\Delta m=\pm 1$. These transitions
are suppressed only as the inverse square of the energy difference
between the initial state and the virtual state involved. In sequential
tunneling, such transitions are also possible, requiring two consecutive steps,
but are \emph{exponentially} suppressed in the Coulomb-blockade regime. In the
sequential-tunneling approximation the molecule would thus
remain in the lowest-energy state with essentially unit
probability. This approximation is evidently invalid for
determining the probabilities in this regime.

Interestingly, the strong effect of cotunneling on the probabilities also leads
to observable effects of \emph{sequential} tunneling on transport in the
cotunneling regime.\cite{LLK04,Koch2} While sequential tunneling starting from
the lowest-energy state is exponentially suppressed, sequential tunneling from
higher-energy states can be possible. With increasing bias voltage, these
higher-energy states become increasingly populated due to \emph{cotunneling},
as Fig.~\ref{Fig2}(b) shows. This leads to sidebands in $dI/dV$ in the
Coulomb-blockade regime that show the linear dependence on the gate voltage
characteristic of sequential tunneling.\cite{Koch2} Strong
electron-phonon coupling can enhance this effect, since it crucially
affects the ratio of the rates for sequential and cotunneling
processes.\cite{Koch2} In our case, these sidebands are very weak, since the
current is controlled by the small cotunneling rates. However, we will
see that the effect on the \textit{probabilities} $P^n$ of molecular
states is significant.

\begin{figure}
\begin{center}
\includegraphics[width=8cm,angle=0]{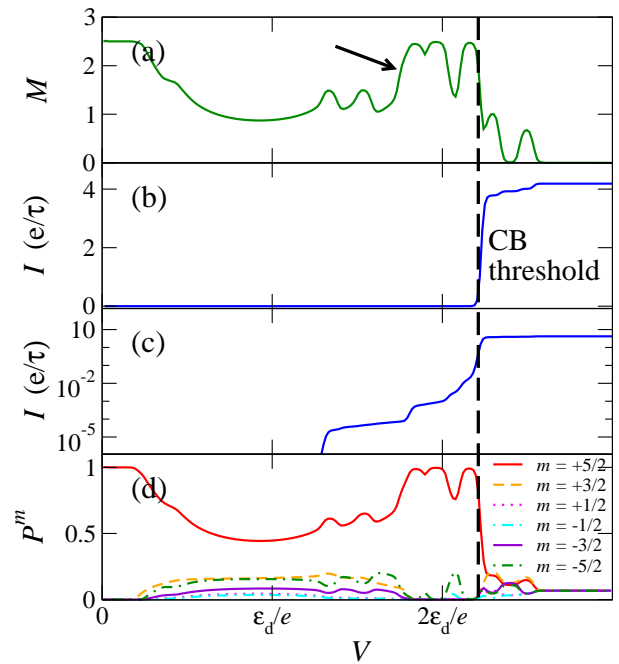}
\caption{(Color online) (a) Magnetization $M$,
(b) linear plot of the current $I$, (c) logarithmic
plot of the sequential-tunneling current, and (d)
probabilities $P^n$ of various molecular many-particle states as functions of
bias voltage $V$.
The parameters are chosen as in Fig.~\protect\ref{Fig2}.}\label{Fig4}
\end{center}
\end{figure}

Figure \ref{Fig4}(a) shows the average magnetization per mo\-le\-cule as a
function of bias voltage over a broad range including both the
cotunneling and sequential-tunneling regimes. The magnetization is nonzero
due to an external magnetic field. At zero bias, the molecule
is in its ground state with $m=5/2$. The onset of inelastic cotunneling to the
two states with $m=3/2$ leads to a decrease in the magnetization in each
case.

The bias-voltage dependence of the magnetization for voltages \emph{above} the
Coulomb blockade  threshold is accompanied by sizeable steps in the current, as
seen in Fig.~\ref{Fig4}(b). At each of these fine-structure steps an additional
inelastic sequential-tunneling transition becomes possible. The
Coulomb-blockade threshold corresponds to the transition with initial state
$n=1$, $m=5/2$ and final state $n=0$, $m=2$. Therefore, the onset of sequential
tunneling is accompanied by a \emph{decrease} in the magnetization. At large
bias the magnetization drops to zero since all states are occupied with equal
probability.

\begin{figure}[t]
\begin{center}
\includegraphics[width=7.5cm]{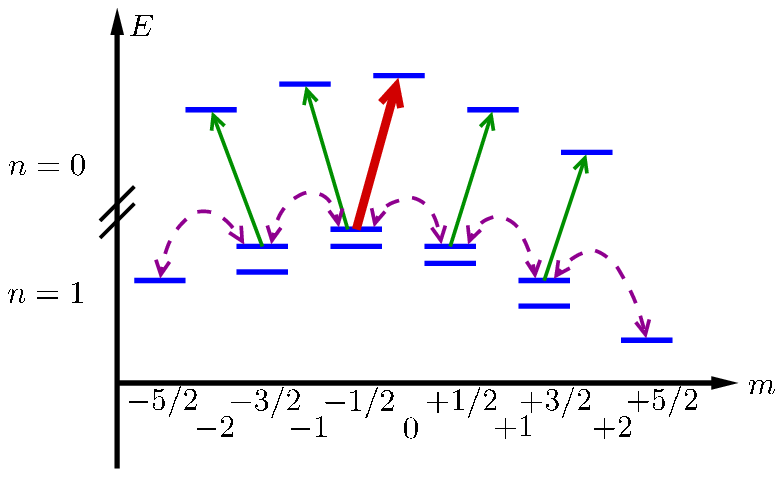}
\caption{(Color online) Level scheme illustrating the interplay between sequential tunneling 
(solid arrows) and cotunneling (dashed arrows)
in magnetic molecules.\protect\cite{footnote.values}
Even below the Coulomb-blockade threshold sequential tunneling
processes starting from higher-energy states populated by cotunneling
may cause the depopulation of these states and drastically affect the 
average magnetization.
At the step denoted by an arrow in Fig.~\protect\ref{Fig4}(a),
the excitation of the transition with $m=-1/2 \to 0$
(heavy solid arrow) gives rise to a redistribution of the 
probabilities $P^n$. Note that exothermal transitions with $\Delta
m=\pm 1/2$ are always possible.}\label{Fig5}
\end{center}
\end{figure}

Remarkably, pronounced step-like features are also present \emph{below} the
Coulomb-blockade threshold in Fig.~\ref{Fig4}(a), where the
current is due to cotunneling and thus very small,
cf.\ Figs.~\ref{Fig4}(b), (c). This can be understood
from the bias-voltage dependence of the relevant
probabilities $P^n$ in Fig.~\ref{Fig4}(d). As an example, consider the step
marked by an arrow in Fig.~\ref{Fig4}(a). The physics leading to the drastic
change of the probabilities is illustrated in Fig.~\ref{Fig5}: The sequential
tunneling processes with $m=-3/2\rightarrow-2$,  $m=-1/2\rightarrow -1$,
$m=3/2\rightarrow 2$, and $m=1/2\rightarrow 1$, starting at the higher-energy
level of each pair (thin arrows in Fig.~\ref{Fig5}), are  already energetically
possible at lower bias voltages causing the partial depopulation of the
initial states. However, the probabilities of these states are non-zero
mainly due to cotunneling processes (dashed arrows in Fig.~\ref{Fig5}). Below
the step marked in Fig.~\ref{Fig4}(a), the half-integer spin states with
positive and negative $m$ are \emph{not connected} by sequential tunneling
processes. As soon as the transition with $m=-1/2\rightarrow 0$ (bold arrow in
Fig.~\ref{Fig5}) becomes possible, the states with positive and negative $m$
are connected and fast, sequential tunneling processes depopulate all states
except for the ground state, which has $m=5/2$. Consequently, the average
magnetization again approaches its maximum value.
Similarly, one can attribute each step to a particular molecular transition.
As Fig.~\ref{Fig4}(c) shows, the onsets of \emph{some} of these
sequential-tunneling processes can also be seen in the sequential-tunneling current, which is,
however, tiny in the cotunneling regime.

The above discussion shows that quantities that depend strongly on the
probabilities of molecular states, such as the magnetization, are much more
sensitive to changes of the bias voltage in the Coulomb-blockade regime than
the conductivity. This suggests to use the \textit{magnetization-voltage}
characteristics, i.e., the magnetization as a function of bias voltage, instead
of the current-voltage characteristics to extract the excitation spectrum  of
magnetic molecules. In order to distinguish magnetic transitions from, e.g.,
vibrational excitations, one should analyze their dependence on the magnetic
field. Furthermore, for a monolayer there is no gate voltage that can serve as
an independent parameter. The magnetic field can assume this role.

\begin{figure}[!t]
\begin{center}
\includegraphics[width=8.0cm]{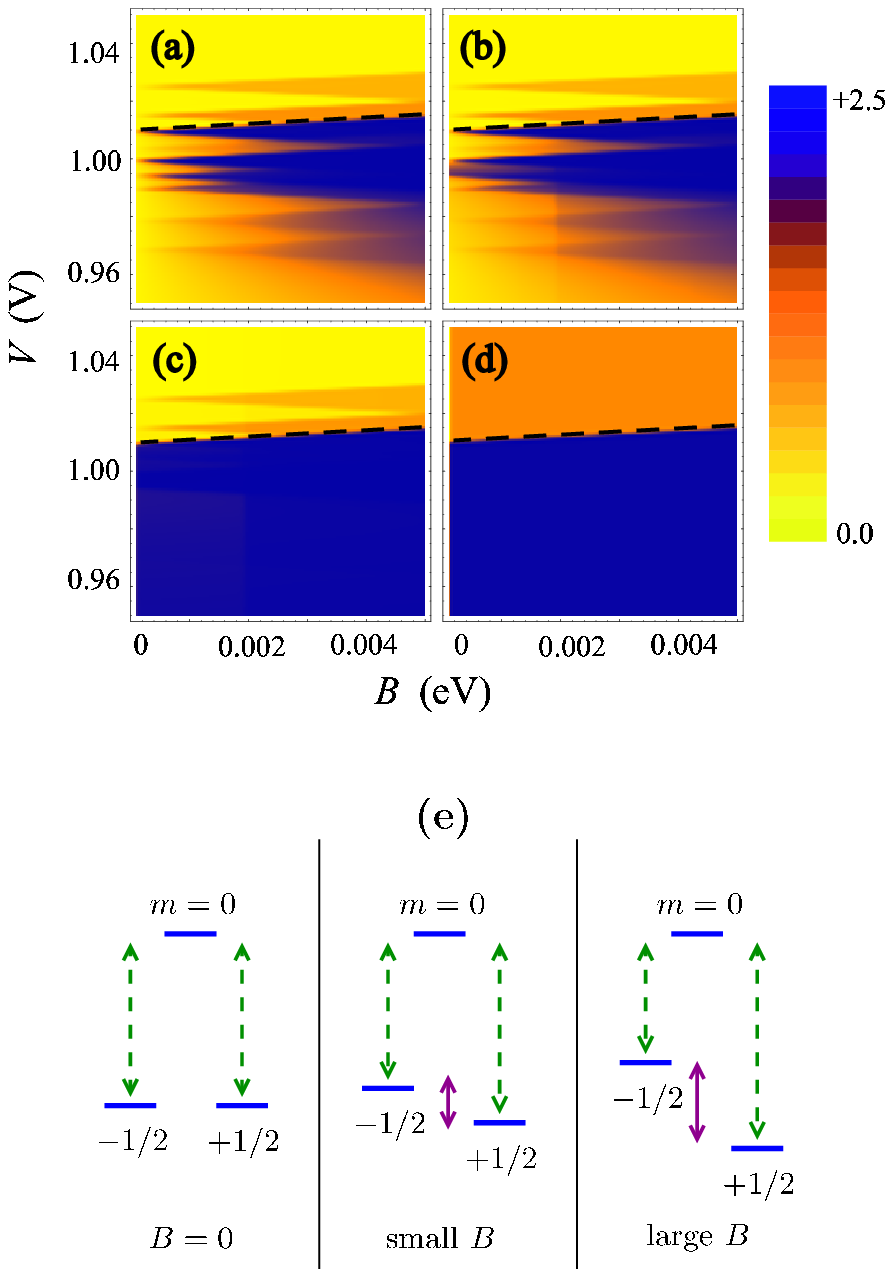}
\caption{(Color online) Magnetization $M$ as a function of bias voltage $V$ and magnetic field $B$
for different spin relaxation times:
(a) $t_{\text{rel}} = \infty$, (b) $t_{\text{rel}} = 10^{6}\tau$,
(c) $t_{\text{rel}} = 10^{2}\tau$, and (d) $t_{\text{rel}} = 0$.
Here $\tau=(2\pi t_\alpha^2 \nu_\alpha)^{-1}$ denotes the typical
electronic tunneling time, assuming symmetric coupling to the leads.
All other parameters are chosen as above.
The dashed lines denote the Coulomb-blockade threshold.
(e) Level schemes illustrating the origin of the magnetization
plateaus.}\label{Fig6}
\end{center}
\end{figure}

Figure \ref{Fig6}(a) shows a density plot of the magnetization as a function of
bias voltage and magnetic field.  The magnetization is an odd function of the
field. The transition energies shift linearly with the field, $\Delta E =
\Delta m\,B$, if the initial and final states have magnetic quantum numbers
differing by $\Delta m$.

Complementary to conventional differential-con\-duc\-tance plots, the density
plots
in Fig.~\ref{Fig6} can serve as fingerprints of the internal degrees of freedom
of the molecules. The Zeeman splitting of the molecular levels due to the
external magnetic field  gives rise to triangular plateaus with a tip at $B=0$.
These plateaus are bounded by two sequential-tunneling transitions. In each
case, these two transitions differ in the sign of the magnetic quantum number
$m$ of both initial and final molecular states. For the chosen parameters, the
plateaus can be attributed to the following transitions from empty to singly
occupied states, starting at low bias voltage (cf.\ Fig.~\ref{Fig5}): $|m|=3/2
\rightarrow 2$, $|m|=1/2\rightarrow 1$, $|m|=1/2\rightarrow 0$,  $|m|=3/2
\rightarrow 2$, $|m|=3/2\rightarrow 1$, $|m|=1/2\rightarrow 1$, 
$|m|=5/2\rightarrow 2$ (this is the first transition starting from the ground
state and thus represents the Coulomb-blockade threshold), $|m|=1/2\rightarrow
0$, and $|m|=3/2\rightarrow 1$. Several transitions appear twice because there
are two states with magnetic quantum numbers $\pm 3/2$ and $\pm 1/2$,
respectively. For a local spin $S=2$ there exist nine transitions obeying the
selection  rule $\Delta m = \pm 1/2$, as can be seen from Fig.~\ref{Fig5}, in
accordance with the nine plateaus shown in Fig.~\ref{Fig6}(a). Note again that
the signal is similar on both sides of the Coulomb-blockade threshold.

The origin of the plateaus is schematically illustrated in
Fig.~\ref{Fig6}(e) for the transition $|m|=1/2\rightarrow 0$. In the absence of
an external Zeeman field the excitation energies for both transitions is equal.
However, the excitation energies differ by the Zeeman energy as soon as a
magnetic field is switched on. This leads to the occurrence of a finite
bias-voltage window, where the excitation of one of the two transitions is
energetically possible whereas the other one is not. Inside this window, only
the spin-down state is depopulated by sequential tunneling, leading to a large
positive magnetization.

So far we have restricted ourselves to the situation where the relaxation of
the local molecular spin is dominated by electron tunneling,
i.e., the spin is \textit{conserved} between tunneling events. However, there
are other processes that also contribute to spin relaxation: (i) Magnetic
molecules containing transition-metal ions, such as $\mathrm{Mn_{12}}$
clusters, show strong spin-orbit interaction, which leads to spin relaxation.
(ii) Hyperfine interactions with nuclear magnetic moments in the
molecule can also lead to spin relaxation. However, in molecules one has the
chance to essentially remove this mechanism by choosing isotopes with vanishing
nuclear spins. (iii) Dipolar interactions with spins of other molecules in the
monolayer or with impurity spins in the electrodes contribute to spin
relaxation. (iv) Small non-uniaxial magnetic anisotropies lead to tunneling
between the eigenstates of $H_{\text{mol}}$. This mechanism has recently been
discussed in the context of transport through magnetic
molecules.\cite{Heersche,Romeike,Romeike2}

All these processes change the magnetic quantum number while keeping the
electron number constant ($\Delta n=0$). The dominant transitions are the ones
with $\Delta m=\pm 1$. These are the same selection rules as for cotunneling,
indicating that one should include additional spin relaxation for consistency
when studying cotunneling.

The effect of spin relaxation on the electronic transport is  included in the
formalism by a phenomenological rate $\propto 1/t_{\text{rel}}$ which forces
the  system to approach the equilibrium distribution on the timescale
$t_{\text{rel}}$. We include additional transition rates between states
$|n\rangle$ and $|m\rangle$ with $\Delta n=0$ and $\Delta m=\pm 1$,
$\Gamma_{nm}^{\text{rel}} = \exp[(\epsilon_n-\epsilon_m)/kT]/t_{\text{rel}}$
for $\epsilon_n < \epsilon_m$ and $\Gamma_{nm}^{\text{rel}} = 1/t_{\text{rel}}$
otherwise. The additional rates obey detailed balance,
ensuring relaxation towards equilibrium in the absence of tunneling.

Effects of spin relaxation on the bias-voltage dependence of the magnetization
are illustrated in Figs.~\ref{Fig6}(a)--(d). For small $t_{\text{rel}}$ (fast
relaxation), the number of  transitions appearing as steps in the
magnetization-voltage characteristics is reduced, since spin relaxation
depopulates higher-energy states that serve as initial states for these
transitions.

The magnetization plateaus start to occur when the relaxation time
$t_\text{rel}$ becomes significantly larger than the typical
sequential-tunneling time $\tau=(2\pi t_\alpha^2\nu_\alpha)^{-1}$. (The
\emph{sequential}-tunneling time enters because the relevant process is
the depopulation of states by sequential tunneling.) Then the time spent by the
electron on the molecule is smaller than the spin relaxation time so that
magnetic excitations survive between tunneling events.

\begin{figure}
\begin{center}
\includegraphics[width=8.0cm]{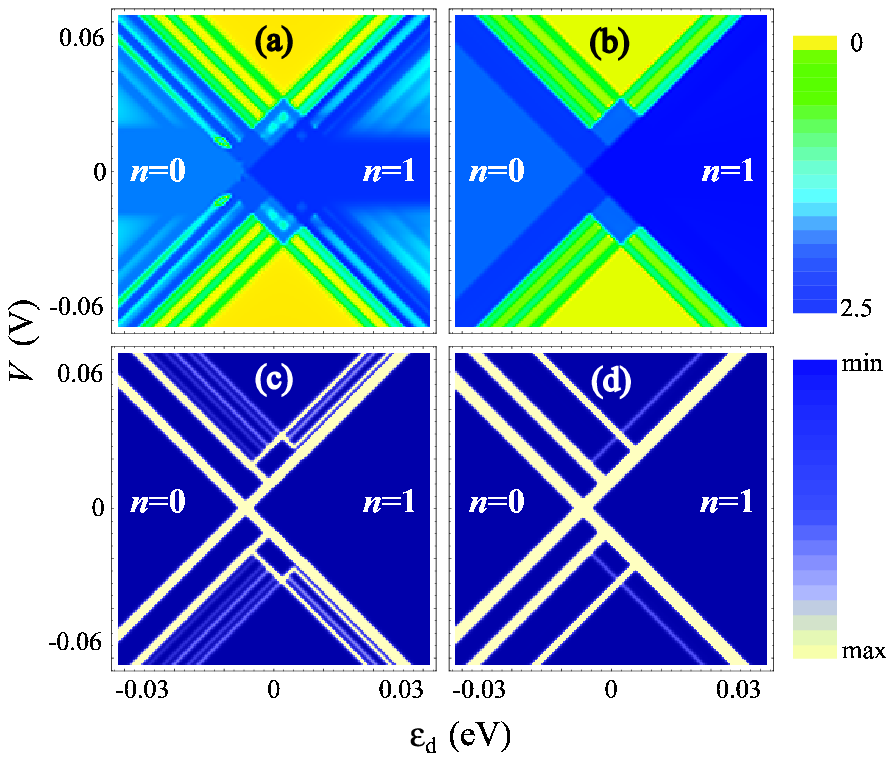}
\caption{(Color online) (a), (b) Magnetization $M$ and (c), (d)
differential conductance
of a single magnetic molecule as a function of $V$ and $\epsilon_d$ for 
(a), (c) slow spin relaxation, $t_{\text{rel}} = 10^{10}\tau$, and
(b), (d) fast spin relaxation, $t_{\text{rel}} = \tau$.
We assume $S=2$, $J=K_2=5\,\mathrm{meV}$, $T=0.1\,\mathrm{meV}$, and
$B=2\,\mathrm{meV}$.}\label{Fig7}
\end{center}
\end{figure}

So far we have considered a monolayer of magnetic molecules, mostly because the
measurement of the magnetization is easier for larger numbers of molecules. As
mentioned previously, even the detection of a \emph{single} molecular spin
might be feasible.\cite{Durkan,Rugar} Using a single molecule allows one to
introduce a gate electrode in order to tune the molecular energy levels by
shifting $\epsilon_d$, see Eq.~(\ref{Hamiltonian}). In the following, we
briefly discuss results obtained for varying gate voltage. To increase the
magnetization signal while retaining the gate electrode, one might consider a
one-dimensional array of magnetic molecules or even a large number of such
arrays aligned in parallel.

The plot of the magnetization and the differential conductance as functions
of bias voltage and onsite energy $\epsilon_d$ presented in
Figs.~\ref{Fig7}(a),(c) shows two striking features. First, the magnetization
shows steps indicating the onset of inelastic  cotunneling which are almost
independent of $\epsilon_d$. The corresponding steps in $dI/dV$ are very
small in absolute units, see Fig.~\ref{Fig2}(a).

Second, the  magnetization shows strong additional magnetic sidebands in the
Coulomb-blockade regime. These sidebands are the consequence of
sequential-tunneling transitions depopulating molecular states that are
populated by cotunneling, as discussed above. In $dI/dV$ the corresponding
features are completely hidden by the low-bias tail of the large peak at the
Coulomb-blockade threshold (not shown). The observation of these sidebands in
the Coulomb-blockade regime  requires spin relaxation times long compared to
the typical tunneling time. For fast spin relaxation, fine-structure peaks are
only present in the sequential-tunneling regime, see Fig.~\ref{Fig7}(b), since
sequential tunneling is still faster than spin relaxation, even though
cotunneling is slower. As shown in Fig.~\ref{Fig7}(d), the absence of such
sidebands is accompanied by suppressed fine-structure peaks in the
sequential-tunneling regime.

\begin{figure}[!t]
\begin{center}
\includegraphics[width=8cm]{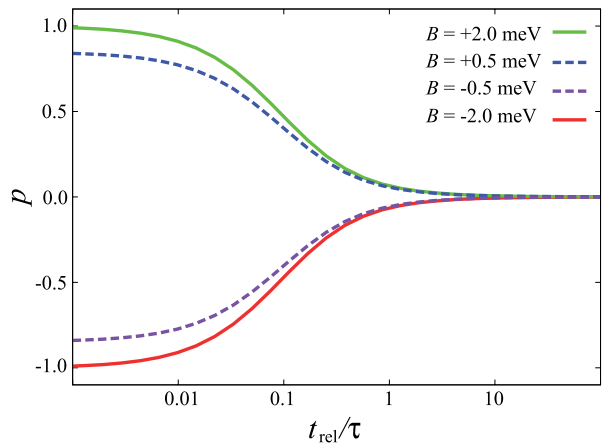}
\caption{(Color online) Polarization of the current,
$p\equiv(I^{\text{L}\uparrow}-I^{\text{L}\downarrow})/
(I^{\text{L}\uparrow}+I^{\text{L}\downarrow})$, 
as a function of spin relaxation time $t_{\text{rel}}$ in units of the
typical tunneling time $\tau$.}\label{Fig8}
\end{center}
\end{figure}

Finally, we note that sufficiently fast spin relaxation leads to
\emph{spin-polarized} stationary currents in the presence of a magnetic field.
If the spin of the magnetic molecule relaxes fast compared to the typical
tunneling rate, which is essentially determined by the current, the system is
always in its ground state. Due to the Zeeman effect the ground
state has maximum magnetic quantum number; $m=5/2$ for our example. Thus only
spin-down electrons can tunnel onto the molecule,  resulting in a
spin-polarized current. Note that this argument is not restricted to
low-order perturbation theory in $H_{\text{t}}$. As shown in Fig.~\ref{Fig8},
the degree of spin polarization is basically determined by the ratio of the
spin relaxation rate and the typical electronic tunneling rate.

\section{Conclusions}

In summary, we have studied the interplay of electronic transport through
magnetic molecules and their non-equilibrium magnetic moment beyond the
sequential-tunneling approximation. We have focused mostly on
monolayers, which should give a better chance to measure the magnetization than
single molecules would.

While the excitation of inelastic tunneling processes in the Coulomb-blockade
regime leads only to a very small absolute change in the current, the change of
the probabilities to find the molecule in various many-particle states is
significant. This manifests itself in a strong bias-voltage dependence of the
magnetization. The magnetization of a molecular monolayer can be switched by an
amount of the order of the saturation magnetization by a small change of bias
voltage, and without causing the flow of a large current.

We find steps in the differential conductance due to inelastic
cotunneling, which have been observed in experiments on
$\mathrm{Mn}_{12}$.\cite{Jo} These steps are accompanied by much larger changes
in the magnetization. Another interesting effect is the
appearance of additional sidebands in the Coulomb-blockade regime that can be
ascribed to \textit{de}-excitations by sequential tunneling of states
populated by cotunneling. These sidebands are very prominent in the
magnetization. We suggest that the magnetization, or any
measurable quantity that strongly differs between molecular states, can be
employed to study molecular transitions that are, from the point of view of
transport, hidden in the Coulomb-blockade regime.

For spintronics applications, the ability to control the persistence of the
stored information is crucial. In this context, we have considered effects of
additional spin relaxation in the same formalism. Our results show that for
sufficiently fast spin relaxation the peaks in the differential conductance and
the steps in the magnetization are washed out, as expected. At the same time,
the degree of polarization of the steady-state current contains information
about the ratio of the spin relaxation rate and the typical electronic
tunneling rate. Fast spin relaxation, while in general undesirable, can lead to
a highly polarized current in the presence of a magnetic field.\\

\acknowledgments

We would like to thank J. Koch and F. von Op\-pen for valuable discussions.
Support by the Deut\-sche For\-schungs\-ge\-mein\-schaft through
Son\-der\-for\-schungs\-be\-reich 658 is gratefully acknowledged.


\begin{thebibliography}{99}
\bibitem{footnote.values}Since we wish to exhibit the general
physical effects expected in (co-) tunneling through monolayers, the parameters
are chosen in such a way that these effects are most clearly seen. The
parameters do not correspond to a specific molecule. Note that a large exchange
interaction, $JS^2\gg K_2S$, which is appropriate for some molecules, does not
change this physics---in Figs.~\ref{Fig3} and \ref{Fig5} it would just shift the
$n=1$ spin quartet to high energies, possibly above the $n=0$ quintet, which
would not interfere with the mechanism of large magnetization
changes.
\end{thebibliography}
\end{document}